\documentclass[aps, prl,10pt,twocolumn,superscriptaddress,nofootinbib]{revtex4-2}
\usepackage[dvipsnames]{xcolor}
\usepackage{mathrsfs, amssymb, amsmath}  
\usepackage{epsfig, cancel}
\usepackage{latexsym}
\usepackage{caption}
\usepackage{natbib, comment}
\usepackage{booktabs}
\usepackage{url}
\usepackage{dcolumn}
\usepackage{multirow}
\usepackage{color}
\usepackage{cancel}
\usepackage{soul}
\usepackage[normalem]{ulem}
\usepackage{amsfonts,amssymb,amsmath, txfonts}
\usepackage{graphicx,epsfig}
\usepackage{psfrag}
\usepackage{hyperref}
\hypersetup{colorlinks=true}
\usepackage{mathtools}
\usepackage{enumitem}
\usepackage{float}
\usepackage[dvipsnames]{xcolor}
\usepackage{xcolor}
\hypersetup{ linktoc=all,
    colorlinks, linkcolor={blue},
    citecolor={darkred}, urlcolor={darkgreen}
}

\usepackage{mathrsfs, amssymb, amsmath, mathtools} 
\usepackage{multirow}
\usepackage{amsfonts,amssymb,amsmath, txfonts}
\usepackage{graphicx,epsfig}
\usepackage{hyperref}
\usepackage{xcolor}
\usepackage{subcaption}
\usepackage{booktabs}
\graphicspath{{Images/}}
\newlength{\DefaultTabColSep}
\definecolor{rosy}{RGB}{230,235,252}
\definecolor{myframetitle}{RGB}{90,89,170}
\definecolor{myblocktitle}{RGB}{140,185,249}
\definecolor{mytitle}{RGB}{10,80,26}

\definecolor{darkgreen}{RGB}{27,130,45}
\definecolor{darkblue}{rgb}{0,0,0.3}
\definecolor{darkred}{rgb}{0.7,0,0}

\definecolor{light gray}{RGB}{220,220,220}
\definecolor{dark purple}{RGB}{108,0,217}
\definecolor{pink}{RGB}{190,20,100}
\definecolor{orang}{RGB}{193,63,0}
\definecolor{green}{RGB}{11,98,17}
\definecolor{darkpink}{RGB}{153,0,76}
\definecolor{bluegreen}{RGB}{0,102,102}
\definecolor{greenlagan}{RGB}{0,102,0}
\definecolor{redgreen}{RGB}{102,102,0}
\definecolor{Redgreen}{RGB}{153,76,0}
\definecolor{vividviolet}{rgb}{0.62, 0.0, 1.0}
\definecolor{amaranth}{rgb}{0.9, 0.17, 0.31}
\definecolor{palatinateblue}{rgb}{0.15, 0.23, 0.89}
\definecolor{brightpink}{rgb}{1.0, 0.0, 0.5}
\definecolor{cornflowerblue}{rgb}{0.39, 0.58, 0.93}
\definecolor{deepcarminepink}{rgb}{0.94, 0.19, 0.22}
\definecolor{radicalred}{rgb}{1.0, 0.21, 0.37}

\def\H0{{\text{H}\hspace*{-2.05mm}\text{H} 0\hspace*{-1.35mm}0\ }}

\def\be{\begin{equation}}
\def\ee{\end{equation}}
\def\beq{\begin{equation}}
\def\eeq{\end{equation}}
\def\bea{\begin{eqnarray}}
\def\eea{\end{eqnarray}}
\newcommand{\dd}{\textrm{d}}

\begin{document}

\title{A Prediction for DESI Full-Shape: Increasing Tomographic \texorpdfstring{$\Omega_m(z)$}{Omegam} Trend}

\author{Eoin \'O Colg\'ain}\email{eoin.ocolgain@atu.ie}
\affiliation{Atlantic Technological University, Ash Lane, Sligo F91 YW50, Ireland}
\author{M. M. Sheikh-Jabbari} \email{jabbari@theory.ipm.ac.ir}
\affiliation{School of Physics, Institute for Research in Fundamental Sciences (IPM), P.O.Box 19395-5531, Tehran, Iran}

\begin{abstract}
To confirm $\Lambda$CDM deviations are due to missing physics (not systematics), one should demonstrate that the model fitting parameters exhibit qualitatively similar redshift drift across independent observables. This is the only way one guarantees new physics. Here, we show that a recent Dark Energy Spectroscopic Instrument (DESI) DR2 Full-Shape (FS) modelling Lyman-$\alpha$ constraint at $z_{\rm eff} = 2.33$ combined with earlier DR1 FS modelling constraints with $0.295 \leq z_{\rm eff} \leq 1.491$ leads to a straight line $\Omega_m(z) = m z + c$ with slope $m = 0.022 \pm 0.012$, $1.8 \sigma$ removed from constant $\Omega_m$. Akaike Information Criterion and Bayesian evidence confirm that constant $\Omega_m$ and increasing $\Omega_m(z)$ are statistically indistinguishable. Through the $Om(z)$ diagnostic, we review how increasing and decreasing $\Omega_m(z)$ trends map to phantom and quintessence dark energy (DE) regimes, respectively. While FS modelling constraints map to phantom DE, the decreasing and increasing $\Omega_m(z)$ trends in DESI BAO and DESI with external data make a phantom crossing inevitable. {Since dynamical DE is but one interpretation for $\Omega_m(z)$ trends, it is imperative that different datasets converge on their $\Omega_m(z)$ trends before one jumps to physical conclusions.} We forecast how DESI FS modelling $\Omega_m$ constraints will improve up to the final data release and explore the implications for model selection.  
\end{abstract}

\maketitle

\section{Introduction}
\vspace*{-2mm}
Systematics aside, the key physical implication of $H_0$ tension \cite{Planck:2018vyg, Riess:2021jrx, H0DN:2025lyy, RoyChoudhury:2026yyk, DiValentino:2021izs, Hu:2023jqc}, and $\Lambda$CDM tensions more generally \cite{CosmoVerseNetwork:2025alb}, is that $\Lambda$CDM model has broken down. Since $\Lambda$CDM is a dynamical model, such an outcome necessitates the fitting parameters of the model, constants defined at redshift $z=0$ (today), pick up redshift dependence \cite{Krishnan:2020vaf}. For the late-Universe background $\Lambda$CDM model, this means $z$-dependence in either of its only two parameters $H_0$ and $\Omega_m$. The only catch is that \textit{a priori} one does not know the redshift range where this happens. This motivates a {tomographic research programme where one bins observables by redshift}, wherever data is available, and ascertains if the fitting parameters evolve. If they do, this is a telltale signature of missing physics that allows to localise the redshift range where physics is missing \cite{Akarsu:2024qiq}. Moreover, by analysing different observables independently, one can in principle identify redshift ranges where different observables fail to track common $\Lambda$CDM physics, thereby suggesting where observational systematics may be at play.   

Unsurprisingly, the claims of redshift dependent $\Lambda$CDM parameters begin with the parameters at the heart of $H_0$ \cite{Planck:2018vyg, Riess:2021jrx, H0DN:2025lyy} (see also \cite{Freedman:2021ahq, Pesce:2020xfe, Kourkchi:2020iyz, Blakeslee:2021rqi}) and $S_8$ tensions \cite{Heymans:2013fya, Joudaki:2016mvz, DES:2017qwj, HSC:2018mrq, KiDS:2020suj, DES:2021wwk}. Observations consistently report a decreasing $H_0$ trend with redshift \cite{Wong:2019kwg, Millon:2019slk, Krishnan:2020obg, Dainotti:2021pqg, Dainotti:2022bzg, Hu:2022kes, Jia:2022ycc, Dainotti:2023yrk, Montani:2023ywn, Mazurenko:2024gwj, DeSimone:2024lvy, Navone:2025gxr, Valletta:2025bgu, Dainotti:2025qxz}, while if $S_8$ evolves, it increases with redshift in the linear regime \cite{Adil:2023jtu, Akarsu:2024hsu}. {Note, CMB lensing restricts any putative $S_8$ tension to $z \lesssim 2$ \cite{ACT:2023kun}, but we caution that by definition $S_8$ must track $\Omega_m$.}
In 2022, invoking a recognised negative correlation between $H_0$ and $\Omega_m$, it was argued, and supporting observations were provided, that $\Omega_m$ increases with redshift \cite{Colgain:2022nlb, Colgain:2022rxy} (see also later \cite{Malekjani:2023ple, Colgain:2024ksa, Colgain:2024clf}).  To the extent of our knowledge, this marked the first claim of a non-constant $\Omega_m$ (an ``$\Omega_m$ tension") predating the DESI collaboration highlighting disagreements in $\Omega_m$ inferences from datasets based on different observables \cite{DESI:2024mwx, DESI:2024hhd, DESI:2024jxi, DESI:2025zgx}. As explained in \cite{Colgain:2024mtg}, $\Omega_m$ differences at different effective redshifts {apparently explain} the DESI collaboration's dynamical dark energy (DE) claim \cite{DESI:2024mwx, DESI:2024hhd, DESI:2024jxi, DESI:2025zgx}.  

To avoid doubt, in this letter the  notation we employ is a deliberate contradiction $\Omega_{m}(z) := \Omega_{m0}(z)$, where we allow integration constants in models the freedom to exhibit redshift dependence. {The analogy is to fit a line to quadratic data, where the slope depends on the range, thus confirming that the line is inadequate.} Following DESI DR1 BAO, it was conjectured in \cite{Colgain:2024xqj} that DESI should converge to an increasing $\Omega_m(z)$ mirroring earlier observations \cite{Colgain:2022nlb, Colgain:2022rxy, Malekjani:2023ple}. We stress that DESI BAO alone and DESI combined with external datasets exhibit more complicated decreasing and increasing $\Omega_m(z)$ trends \cite{Colgain:2024mtg} which, as we shall argue, if interpreted as dynamical DE, make phantom crossing DE inevitable. 

Here, we recall that observations based on combined observables are weaker due to a greater set of potential systematics. Indeed, SNe systematics are a concern \cite{Efstathiou:2024xcq, Gialamas:2024lyw, Dhawan:2024gqy, Afroz:2025iwo}. Moreover, comparisons of DESI Full-Shape (FS) modelling and DESI BAO reveal that tomographic constraints on the $\Lambda$CDM parameter $\Omega_m$ have yet to fully converge \cite{Colgain:2025nzf, Colgain:2025fct}. Nevertheless, FS modelling uses more information in the data, so one expects FS modelling to be more robust than BAO to statistical fluctuations. The key point of this letter is that the increasing $\Omega_m(z)$ trend in DESI data conjectured in \cite{Colgain:2024xqj} may be emerging in DESI FS modelling constraints. Interestingly, an increasing $\Omega_m(z)$ trend maps to phantom dynamical DE (no crossing). This has the upshot that since $w(z)$, in particular $w_0 := w(z=0)$, and $H_0$ are typically negatively correlation \cite{Vagnozzi:2018jhn, Vagnozzi:2019ezj, Alestas:2020mvb, Lee:2022cyh}, phantom DE is less at odds with $H_0$ tension. Note, the spirit of this letter is not to jump to conclusions on dynamical DE, rather to address a prerequisite question: Is $\Lambda$CDM breaking down through $\Omega_m$ parameter drift?

\vspace*{-4mm}
\section{Increasing $\Omega_m(z)$}
\vspace*{-4mm}

\begin{figure}[H]
\centering
\includegraphics[scale=0.5]{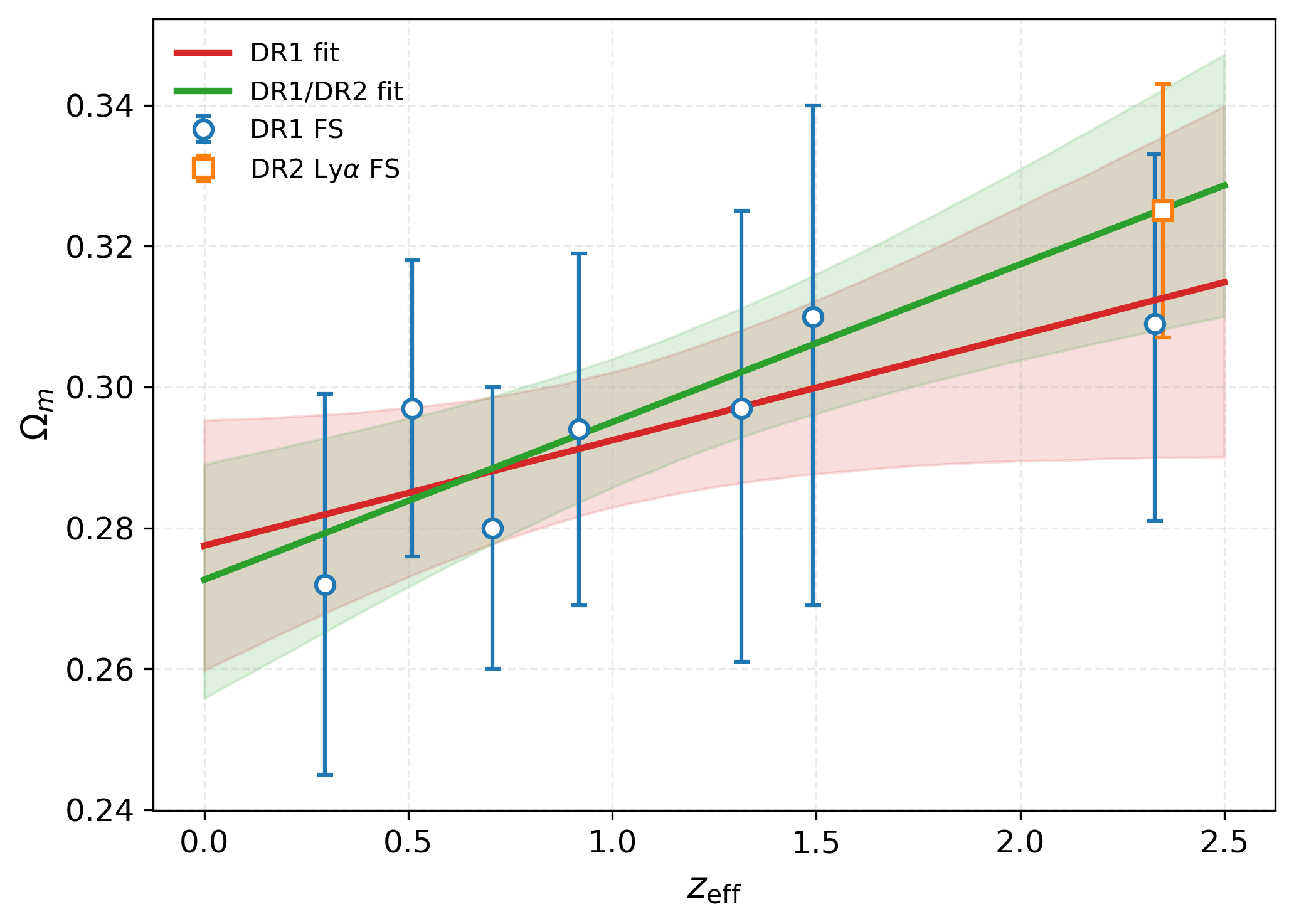}
\caption{Increasing $\Omega_m (z)$ trend with DR1 FS modelling $\Omega_m$ constraints and mixed DR1/DR2 $\Omega_m$ constraints with statistical signficance $1 \sigma$ and $1.8 \sigma$, respectively.}
\label{fig:dr1_dr2}
\end{figure}

We begin by fitting a line model 
\begin{equation}
\label{eq:line}
\Omega_m (z) = m z + c
\end{equation}
 to the DR1 FS $\Omega_m$ constraints \cite{DESI:2024jxi} in Table \ref{tab:FS} without the DR1 Lyman-$\alpha$ constraint \cite{Cuceu:2025nvl}. For simplicity, we symmetrise the errors and find $m = 0.021 \pm 0.029$ corresponding to a significance of $0.7 \sigma$. We next include the DR1 Lyman-$\alpha$ constraint, which increases the  significance to $1 \sigma$, $m = 0.015 \pm 0.015$. Finally, we replace the DR1 Lyman-$\alpha$ constraint with the DR2 Lyman-$\alpha$ constraint \cite{DESI:2026lnd} finding $m = 0.022 \pm 0.012$ with significance $1.8 \sigma$. We illustrate the best fit lines and $68 \%$ credible intervals for DR1 constraints and mixed DR1/DR2 constraints in Fig. \ref{fig:dr1_dr2}. {While the increasing $\Omega_m(z)$ trend is evident to the eye in DR1 data in Fig. \ref{fig:dr1_dr2}, it is no secret that the recent result \cite{DESI:2026lnd} strengthens the trend.}

\begin{table}[t]
\centering
\renewcommand{\arraystretch}{1.3}
\begin{tabular}{|c|c|c|}
\hline
Tracer & $z_{\rm eff}$  & $\Omega_m$ \\
\hline\hline
BGS  & $0.295$ & $0.272 \pm 0.027$ \\
LRG1  & $0.510$ & $0.297 \pm 0.021$ \\
LRG2  & $0.706$ & $0.280 \pm 0.020$ \\
LRG3  & $0.919$ & $0.294 \pm 0.025$ \\
ELG2  & $1.317$ & $0.297^{+0.028}_{-0.036}$ \\
QSO  & $1.491$ & $0.310^{+0.030}_{-0.041}$ \\
Lyman-$\alpha$  & $2.33$ & $0.309^{+0.024}_{-0.028}$ \\
Lyman-$\alpha$ (DR2)  & $2.33$ & $0.325 \pm 0.018
$ \\
\hline
\end{tabular}
\caption{DESI DR1 $\Omega_m$ constraints with Lyman-$\alpha$ upgraded to a DR2 constraint.}
\label{tab:FS}
\end{table}

We next turn our attention to whether the mixed DR1/DR2 FS constraints prefer the line model (\ref{eq:line}) over a constant $\Omega_m(z) = c$. We begin with the Akaike Information Criterion (AIC) \cite{Akaike:1974vps}, where we consider both the base AIC and its correction for small samples:
\begin{equation}
\mathrm{AIC} = \chi^2_{\rm min} + 2 k, \qquad 
\mathrm{AIC}_c =  \chi^2_{\rm min}+ \frac{2 k N }{N-k-1} 
\end{equation}
where $\chi^2_{\rm min}$ is the minimum of the chi-square, $k$ is the number of free parameters and $N = 7$ is the size of the sample. Comparing the line with two parameters to a constant horizontal with one parameter, we find $\Delta \chi^2_{\rm min} = -3.5$ for the more complex model leading to $\Delta \mathrm{AIC} = -1.4$ but $\Delta \textrm{AIC}_c = 0.7$. Despite the difference in sign between AIC and $\textrm{AIC}_c$, $|\Delta \rm {AIC}| \lesssim 2$ confirms that the two models are statistically indistinguishable given current data. 

We can also calculate Bayesian evidence $Z$ to get a second perspective. Given the simplicity of the models and likelihood, it is easy to calculate 
\begin{equation}
\begin{aligned}
Z_0 &= \int_{c=0.2}^{c=0.4} e^{-\frac{1}{2} \chi^2(c)} \, \pi(c) \, \dd c, \\
Z_1 &= \int_{m = -0.04}^{m = 0.07} \int_{c = 0.2}^{c=0.4} e^{-\frac{1}{2} \chi^2(m, c)} \, \pi(m) \, \pi(c) \, \dd m \, \dd c
\end{aligned}
\end{equation}
where we have chosen uniform priors $\pi(x) = 1/(x_{\rm max}-x_{\rm min})$ so that they are minimal and never truncate the posterior. We find $B_{10} = Z_1/Z_0 = 1.5$, which shows weak preference for the line. Nevertheless, what the ${\rm AIC}$ and $B_{10}$ result tell us is that current data does not strongly distinguish the models.

It cannot be stressed enough what this means. In cosmology model parameters such as $\Omega_m$ are constant by assumption. However, a physical $H_0$ tension necessitates an evolving $\Lambda$CDM fitting parameter $H_0$ at some redshift \cite{Krishnan:2020vaf} and descending trends at lower redshifts have been observed \cite{Wong:2019kwg, Millon:2019slk, Krishnan:2020obg, Dainotti:2021pqg, Dainotti:2022bzg, Hu:2022kes, Jia:2022ycc, Dainotti:2023yrk, Montani:2023ywn, Mazurenko:2024gwj, DeSimone:2024lvy, Navone:2025gxr, Valletta:2025bgu, Dainotti:2025qxz}. In the $\Lambda$CDM model, $\Omega_m$ is negatively correlated with $H_0$, so this implies an increasing $\Omega_m$ trend, first reported in 2022 \cite{Colgain:2022nlb, Colgain:2022rxy}. If physical, one eventually reaches a point where model selection fails to distinguish between a constant and a redshift dependent evolving $\Omega_m(z)$ before data improves further and the latter is preferred. We may be at the crossover point. We now turn attention to how this overlaps with the DESI collaboration's dynamical DE claim from 2024 onwards \cite{DESI:2024mwx, DESI:2024hhd, DESI:2024jxi, DESI:2025zgx} (see \cite{DES:2024jxu, Rubin:2023ovl} for earlier claims).  

\vspace*{-5mm}
\section{Dynamical Dark Energy Interpretation}
\vspace*{-4mm}

One may map the increasing $\Omega_m(z)$ trend in the $\Lambda$CDM  from Fig. \ref{fig:dr1_dr2} into the $w_0 w_a$CDM (CPL) model \cite{Chevallier:2000qy, Linder:2002et}. The reverse operation, namely mapping a $w_0 w_a$CDM model preferred by DESI \textit{with external datasets} has been performed in \cite{Colgain:2024mtg}. Therein, noting the hierarchies in matter density $\Omega_{m}^{\rm DESI} < \Omega_m^{\rm CMB} < \Omega_m^{\rm SNe}$ and effective redshifts $z_{\rm{eff}}^{\rm SNe} < z_{\rm{eff}}^{\rm DESI} < z_{\rm{eff}}^{\rm CMB}$, both increasing and decreasing $\Omega_m (z)$ behaviours were observed. From lower redshifts, characterised by $z_{\rm{eff}}^{\rm SNe}$, to intermediate redshifts, characterised by $z^{\rm DESI}_{\rm eff}$, $\Omega_m (z)$ decreases sharply but increases again gradually from intermediate redshifts to $z_{\rm{eff}}^{\rm CMB}$ \cite{Colgain:2024mtg}. The behaviour simply reflects the above hierarchies.  

Remarkably, the $Om(z)$ diagnostic \cite{Sahni:2008xx} provides a valuable insight into the trends noted in \cite{Colgain:2024mtg}. Recall its definition 
\begin{equation}
\label{eq:om}
Om(z) \equiv \frac{E(z)^2-1}{(1+z)^3-1} = \Omega_m + (1-\Omega_m) \frac{X(z)-1}{(1+z)^3-1}, 
\end{equation}
where $E(z) := H(z)/H_0$ denotes the normalised Hubble parameter and the second equality is written for a $w(z)$CDM model with,    
\begin{equation}\label{eq:X}
X(z)= \exp \left(3 \int_0^z \frac{1+w(z^{\prime})}{1+z^{\prime}} \dd z^{\prime} \right).   
\end{equation}
$Om(z)$ comprises a constant $\Omega_m$ and a redshift dependent component that under the traditional assumption that DE is irrelevant at higher redshifts leads to $Om(z \rightarrow \infty) = \Omega_m$.\footnote{Here, one only needs to worry about $1+w(z) > 0 \Rightarrow X(z) > 1$ at higher redshifts. In the CPL model, $X(z) \sim (1+z)^{3 (1+w_0+w_a)}$ for $z \gg 1$, so this explains why one typically imposes $w_0 + w_a < 0$ \cite{DESI:2024mwx} to ensure that $(1+z)^3$ dominates $X(z)$.} 
We remark that since $z$ is observationally better constrained than $X(z)$, one typically expects $Om(z)$ to blow up at the lowest redshifts, instead of recovering the ``theoretical'' $\Omega_m$ value; $Om(z)$ is designed to work well in intermediate redshifts, $0.1\lesssim z\lesssim 4-5$, where DE is expected to be relevant and in general it loses sensitivity at lower $z$. For quintessence DE models $X(z)>1$, thus $Om(z)>\Omega_m$ \textit{decreases} with redshift, while for phantom DE models $X(z) < 1$, thus $Om(z)<\Omega_m$ \textit{increases} with redshift. All of this is dictated by the sign of $1+w(z)$. 

{We can now explain why DESI reports a phantom crossing. Provided one uses combined datasets with hierarchies $\Omega_{m}^{\rm DESI} < \Omega_m^{\rm CMB} < \Omega_m^{\rm SNe}$ in the $\Lambda$CDM model and effective redshifts $z_{\rm{eff}}^{\rm SNe} < z_{\rm{eff}}^{\rm DESI} < z_{\rm{eff}}^{\rm CMB}$, $\Omega_m$ decreases with redshift before increasing again \cite{Colgain:2024mtg}. The only way one can exhibit these trends is if $1+w(z)$ changes sign. Note, this cannot happen in the $w$CDM model with constant $w$; the model can only model a decreasing or increasing trend, so CPL is the next simplest choice. Groups have studied the robustness of the phantom crossing \cite{DESI:2024kob, Giare:2024gpk}, but viewed through the prism of $\Omega_m(z)$, it is inevitable if the current $\Omega_m$ hierarchies in combined datasets persist and the only assumptions are pressureless matter and dynamical DE. Note, qualitatively similar hierarchies persist with DESI BAO alone \cite{Colgain:2024xqj, Colgain:2025nzf}, and no surprise, the $w_0 > -1, w_a < 0$ quadrant is preferred.}


\begin{table}[t]
\centering
\renewcommand{\arraystretch}{1.3}
\begin{tabular}{|c|c|c|c|}
\hline
Model & $\Omega_m$  & $w_0$ & $w_a$ \\
\hline\hline
$w$CDM  & $0.325^{+0.016}_{-0.017}$ & $-1.130^{+0.075}_{-0.081}$ & $-$  \\
$w_0 w_a$CDM  & $0.336^{+0.019}_{-0.026}$ & $-1.04^{+0.17}_{-0.14}$ & $-0.8^{+1.2}_{-1.3}$  \\
\hline
\end{tabular}
\caption{$68 \%$ credible intervals for $Om(z)$ models based on $w$CDM and $w_0 w_a$CDM fitted to DR1/DR2 DESI FS constraints on $\Omega_m$.}
\label{tab:om_constraints}
\end{table}

\begin{figure}[t]
\centering
\includegraphics[scale=0.5]{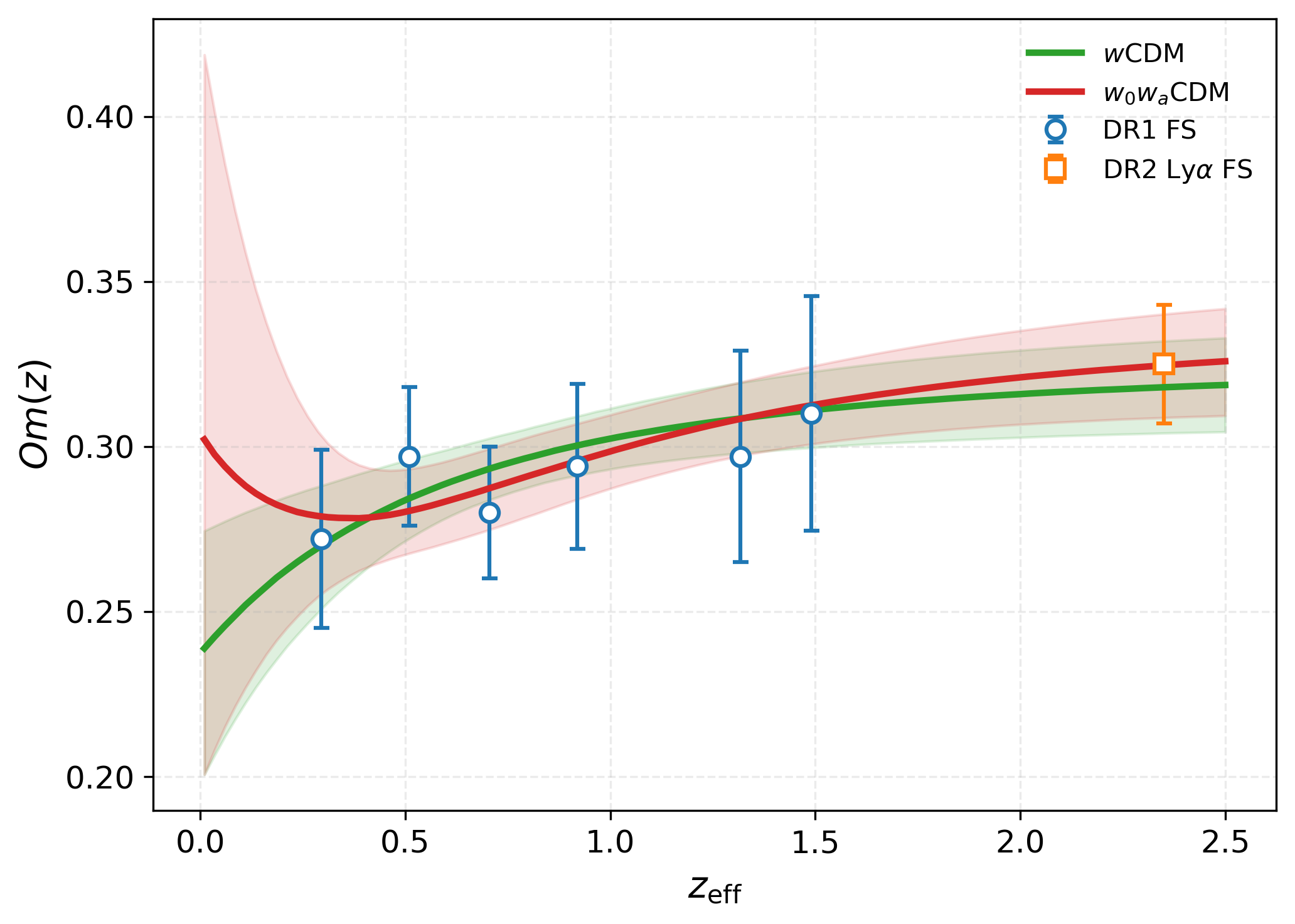}
\caption{$68 \%$ $Om(z)$ credible intervals fitted to DR1/DR2 DESI FS constraints on $\Omega_m$.}
\label{fig:om}
\end{figure}

We stress that in our context here we only consider DESI FS modelling where $\Omega_m$ values show an apparent increasing $\Omega_m(z)$ trend. Therefore, we expect to recover phantom dynamical DE. 
We report $68 \%$ credible intervals for the parameters ($\Omega_m, w_0, w_a)$ in Table \ref{tab:om_constraints} where it should be noted that we treat $Om(z)$ as a model fitted to data in Table \ref{tab:FS}. Evidently, the $\Omega_m$ constraint expectedly tracks the Lyman-$\alpha$ value $\Omega_m \sim 0.325$ (Table \ref{tab:FS}) and both models show the anticipated preference for the $w(z) < 1$ regime. For the $w$CDM model we find $w < -1$ at $ 1.7 \sigma$, but for $w_0 w_a$CDM, while the posteriors inhabit the expected quadrant, the englarged parameter space means that $\Lambda$CDM behaviour is recovered within $1 \sigma$. In Fig. \ref{fig:om} we show the corresponding reconstructed $Om(z)$ credible intervals. We remind the reader that as discussed above it is difficult to read into $Om(z)$ behaviour extrapolated to $z=0$.

\vspace*{-4mm}
\section{Forecast}
\vspace*{-4mm}

Our analysis to this point is based on existing data. In this section we attempt to forecast how the statistical significance of the increasing $\Omega_m(z)$ trend may increase with possible future data. We anticipate that DESI will release updated DR1 $\rightarrow$ DR2 $\Omega_m$ constraints for FS modelling. We note that the Lyman-$\alpha$ upgrade has led to (symmetrised) errors $\pm 0.026$ reducing to $\pm 0.018$, which is an approximate $1/\sqrt{2}$ reduction in error. The simplest DR2 forecast one can make is to assume that the central values do not change, $\Omega_m^{\rm DR2} = \Omega_m^{\rm DR1}$ at $ 0.295 \leq z_{\rm eff} \leq 1.491$, but the DR1 errors reduce $\sigma_{\Omega^{\rm DR2}_m}  = \sigma_{\Omega^{\rm DR1}_m}/\sqrt{2}$. Under this assumption, we find the significance of the slope $m = 0.022 \pm 0.011$ increases to $2 \sigma$. Note, the slope cannot change as we have not shifted the central values. Following this through to Bayesian model comparison, this would lead to  $\Delta \chi^2_{\rm min} = -4.8$, implying $\Delta \mathrm{AIC} = -2.8$,  $\Delta \mathrm{AIC}_c = -0.6$ and $B_{10} = 2.2$, where it should now be noted that there is no longer ambiguity and all model-selection statistics consistently, but weakly, point to the line model. 

Admittedly, the assumption that the DR2 central $\Omega_m$ values stay the same is unrealistic. To improve upon this we take the Lyman-$\alpha$ DR2 $\Omega_m$ constraint \cite{DESI:2026lnd} at face value (no change), but draw new DR2 $\Omega_m$ values in a normal distribution $\Omega_m^{\rm DR2} = \mathcal{N} (\Omega_m^{\rm DR1}, \sigma_{\Omega_m^{\rm DR2}}^2)$ for $0.295 \leq z_{\rm eff} \leq 1.491$ with $\sigma_{\Omega_m^{\rm DR2}}$ defined as above. We run $10^4$ mocks and for each mock we calculate the slope significance $S = m/\sigma_{m}$, $\Delta \mathrm{AIC}$, $\Delta \rm{AIC}_c$ and Bayesian evidence $B_{10}$. We find $S = 2.1\pm 0.6 \sigma$, $\Delta \mathrm{AIC} = -2.5^{+2.2}_{-2.9}$, $\mathrm{AIC}_{c} = -0.2^{+2.2}_{-2.9}$ and $B_{10} = 2.2^{+6.8}_{-1.4}$, where we quote median ($50^{\mathrm{th}}$ percentile) central values and $68 \%$ intervals at $16^{\mathrm{th}}$ and $84^{\mathrm{th}}$ percentiles. The lopsided errors for $B_{10}$ mean that $40 \%$ of our mocks inhabit $B_{10} >3$ and $14 \%$ inhabit $B_{10} > 10$, corresponding to moderate and strong evidence on the Jeffreys' scale for an increasing $\Omega_m(z)$. 

To provide a final data release forecast for DESI, we adopt the latest DR1/DR2 $\Omega_m$ central values and their DR1 symmetrised errors. As we have seen, Lyman-$\alpha$ errors between DR1 \cite{Cuceu:2025nvl} and DR2 \cite{DESI:2026lnd} led to a $1/\sqrt{2}$ reduction. For the final DESI data release, we assume that the final errors are half the original DR1 errors. Thus, relative to DR2, we are assuming a further $1/\sqrt{2}$ reduction. In contrast to our DR2 mocks, where we kept Lyman-$\alpha$ fixed, we allow all $\Omega_m$ values to be drawn from a normal distribution $\Omega_m^{\rm final} = \mathcal{N} (\Omega_m^{\rm DR1/DR2}, \sigma_{\Omega_m^{\rm final}}^2)$, where $\sigma_{\Omega_m^{\rm final}} = \sigma_{\Omega_m^{\rm DR1}}/2$. Running $10^4$ mocks, one forecasts the statistical significance of the slope to increase to $S = 2.9 \pm 1.0 \sigma$, while $\Delta \mathrm{AIC}$ and $\Delta \mathrm{AIC}_c$ should reduce to $-6.6^{+4.9}_{-6.8}$ and $-4.4^{+4.9}_{-6.8}$, respectively, thus providing moderate support for the increasing $\Omega_m(z)$. In addition, we forecast Bayesian evidence in the range $B_{10} = 13^{+369}_{-12}$ with $70 \%$ and $53 \%$ of our mocks in the range $B_{10} > 3$ and $B_{10} > 10$ correspond to moderate and strong support for increasing $\Omega_m(z)$, respectively. The main takeaway lesson is that a strong statement on model selection may have to wait until final DESI results. {It should be stressed that since the shifts in $\Omega_m$ central values with $z$ are not large, the errors play a key role.}  

\vspace*{-5mm}
\section{Conclusions}
\vspace*{-4mm}
Cosmology traditionally makes progress through Bayesian model comparison where the fitting parameters in each model are assumed to be constant. While model comparisons have limitations, including prior dependence, parameterisation sensitivity and the risk of being overly conservative \cite{Liddle:2007fy, Trotta:2008qt, Amendola:2024prl}, arguably the greater weakness is that \textit{model comparison is not a consistency check} and observational systematics can propagate unchecked to mimic physics. More generally, one rarely performs consistency checks of a model, e. g. $\Lambda$CDM, fitted to a dataset to test the constant fitting parameters assumption \cite{Akarsu:2024qiq}. 

Focussing on DESI FS modelling constraints on the $\Lambda$CDM parameter $\Omega_m$ \cite{DESI:2024jxi, Cuceu:2025nvl, DESI:2026lnd}, which exhibit greater convergence towards constant $\Omega_m$ compared to DESI BAO \cite{DESI:2024mwx, DESI:2025zgx} (Fig. 5 of \cite{Colgain:2025nzf}), our analysis shows that it is no longer clear that a constant $\Omega_m$ model is preferred over an $\Omega_m(z)$ model (\ref{eq:line}) tracing a line. This outcome is expected at some point in time if the $\Lambda$CDM model has broken down and $\Lambda$CDM tensions are not due to systematics. {Note that with a single observable, i.e. FS modelling, one can never preclude a systematic explanation, but establishing this is ultimately a matter for the DESI collaboration. What helps mitigate systematics is the independent appearance of qualitatively similar
increasing $\Omega_m(z)$ trends in different observables. In particular, increasing $\Omega_m(z)$ trends have previously been reported in SNe \cite{Colgain:2022nlb,Colgain:2022rxy,Malekjani:2023ple,Colgain:2024ksa}.
Should DESI exhibit the same behaviour, the case for survey-specific systematics would become difficult to sustain.}

Exploiting the $Om(z)$ diagnostic \cite{Sahni:2008xx}, one approach of mapping a normalised Hubble parameter $E(z)$ back to the $\Lambda$CDM parameter $\Omega_m$ \textit{cf.} \cite{Colgain:2024mtg}, we have reviewed how increasing and decreasing $\Omega_m(z)$ trends map to phantom $w(z) < -1$ and quintessence $w(z) > -1$ dynamical DE regimes, respectively. This explains how DESI ends up in the Quintom quadrant with $w_0 > -1, w_a < 0$. When datasets are combined, SNe samples \cite{Brout:2022vxf, Rubin:2023ovl, DES:2024jxu, DES:2025sig} at lower redshifts and CMB \cite{Planck:2018vyg, SPT-3G:2025bzu, ACT:2025blo} at higher redshifts prefer larger $\Omega_m$ values relative to DESI BAO/FS modelling \cite{DESI:2024mwx, DESI:2024hhd, DESI:2024jxi, DESI:2025zgx} at intermediate reshifts. Moreover, DESI BAO alone exhibit apparent fluctuations with the same qualitative trend \cite{Colgain:2024xqj, Colgain:2025nzf}, which leads one again to the quadrant $w_0 > -1, w_a < 0$. The key point is that a decreasing $\Omega_m(z)$ followed by an increasing $\Omega_m(z)$ \cite{Colgain:2024mtg} necessitates a transition from quintessence to phantom DE regimes. Evidently, in DESI FS modelling alone, only an increasing $\Omega_m(z)$ trend is present, so one expects phantom dynamical DE $(w_0 < -1, w_a <0)$. Our $Om(z)$ analysis confirms this. We note that Hubble tension also prefers more negative $1+w(z)$ values \cite{Vagnozzi:2018jhn, Vagnozzi:2019ezj, Alestas:2020mvb, Lee:2022cyh}.

Given that current data where DR1 FS modelling $0.295 \leq z_{\rm eff} \leq 1.491$ \cite{DESI:2024jxi} and DR2 Lyman-$\alpha$ FS modelling at $z_{\rm eff} = 2.33$ \cite{DESI:2026lnd} fail to distinguish a constant $\Omega_m$ from an increasing $\Omega_m(z)$ in model selection statistics, we provide forecasts for the next DR2 upgrade and the final DESI result. As errors contract, our mock forecasts confirm that the statistical significance of increasing $\Omega_m(z)$ should increase to $\sim 2 \sigma$ with the imminent DR2 upgrade and $\sim 3 \sigma$ in the final data release. Although the DR2 upgrade is unlikely to lead to a conclusive result, the final data release may lead to moderate to strong support based on AIC and Bayesian evidence for the line model provided trends persist.

\vspace*{-5mm}
\section*{Acknowledgements} 
\vspace*{-4mm}
This article/publication is based upon work from COST Action CA21136 – “Addressing observational tensions in cosmology with systematics and fundamental physics (CosmoVerse)”, supported by COST (European Cooperation in Science and Technology). The work of MMShJ is in part supported by the INSF research chair grant No.40405163. 

\bibliography{refs}

\end{document}